\documentclass[aip,jmp,superscriptaddress,preprint,author-year]{revtex4-1}

\usepackage{settings}  % External packages and setting should go into settings
\usepackage{macros}    % User-defined commands and envs. should go into macros.

\graphicspath{{figures/}}

\begin{document}

	\title{An Interactive Approach for Identifying Structure Definitions}
	
	\author{Natalia Mikula}
	\affiliation{Zuse Institute Berlin, Visual and Data-centric Computing, 14195 Berlin, Germany}
	
	\author{Tom Dörffel}
	\affiliation{Department of Mathematics and Computer Science, Freie Universität Berlin, 14195 Berlin, Germany}
	
	\author{Daniel Baum}
	\affiliation{Zuse Institute Berlin, Visual and Data-centric Computing, 14195 Berlin, Germany}
	\email{\{mikula, baum\}@zib.de}
	
	\author{Hans-Christian Hege}
	\affiliation{Zuse Institute Berlin, Visual and Data-centric Computing, 14195 Berlin, Germany}
	
	\date{Received: date / Accepted: date}
	% The correct dates will be entered by the editor

        %%%%%%%%%%%%%%%%%%%%%%%%%%%%%%%%%%%%%%%%%%%%%%%%%% 
	
        %-------------------------------------------------------------------------
	\begin{abstract}
		Our ability to grasp and understand complex phenomena is essentially based on recognizing structures and relating these to each other.
		For example, any meteorological description of a weather condition and explanation of its evolution recurs to meteorological structures, such as convection and circulation structures, cloud fields and rain fronts. 
		All of these are spatiotemporal structures, defined by time-dependent patterns in the underlying fields.
		Typically, such a structure is defined by a verbal description that corresponds to the more or less uniform, often somewhat vague mental images of the experts.
		
		%%% 
		However, a precise, formal definition of the structures or, more generally, concepts is often desirable, e.g., to enable automated data analysis or the development of phenomenological models.
		Here, we present a systematic approach and an interactive tool to obtain formal definitions of spatiotemporal structures.
		The tool enables experts to evaluate and compare different structure definitions on the basis of data sets with time-dependent fields that contain the respective structure. 
		Since structure definitions are typically parameterized, an essential part is to identify parameter ranges that lead to desired structures in all time steps.
		In addition, it is important to allow a quantitative assessment of the resulting structures simultaneously.
		We demonstrate the use of the tool by applying it to two meteorological examples: finding structure definitions for vortex cores and center lines of temporarily evolving tropical cyclones.
		
		%%% 
		Ideally, structure definitions should be objective and applicable to as many data sets as possible.
		However, finding such definitions, e.g., for the common atmospheric structures in meteorology, can only be a long-term goal.
		The proposed procedure, together with the presented tool, is just a first systematic approach aiming at facilitating this long and arduous way.

		
	\end{abstract}  
	%-------------------------------------------------------------------------
	
	\maketitle
	
	\section{Introduction}
	
	In order to understand a phenomenon hidden in data, we need to recognize structures and relationships between them.
	Structures are characterized by defining qualities and features.
	In this work we deal with structures for which a mental image does exist but no concise mathematical definition.
	For the sake of clarity, we distinguish between the semantics of a concept, i.e., anything that defines and characterizes it, and the linguistic term used to denote it.
	
	In all areas of daily life as well as in all sciences, concepts are needed for thinking and communicating.
	This can be explained very nicely using the example of meteorology:
	The description of a weather situation and its development over time makes use of technical terms such as ``pressure systems'' (high or low), ``fronts'' (warm, cold or rain), ``clouds'', ``precipitation'', ``jet streams'' and so on.
	Of course, in technical language, much more differentiating terms are used.
	For example, there is a hierarchical classification system for clouds that distinguishes more than 30 types of clouds by classifying them according to altitude and vertical shape, instability or convection activity, and specific structural details~\citep{wmo_cloud}.
	
	%%%
	One could counter that in strictly rational science such mental constructs are dispensable; for example, even a very complex meteorological state is \emph{completely} described by a set of fields, which is governed by a set of equations. One could claim that given these fields in appropriate resolution, there is nothing more to say. What this counterargument overlooks is that humans need both categorizing and characterizing concepts of structures, and associated linguistic terms, in order to think and talk about the phenomena as well as to formulate hypotheses and theories about them. Therefore, such mental constructs are fundamental building blocks of human knowledge and are used practically everywhere.
	
	%%%
	It is an essential part of science to sharpen concepts, to differentiate them more precisely, and to categorize them more meaningfully. Yet, for most concepts, only verbal descriptions exist that are not very precise. Just consider questions like these: ``Are we in the fourth \emph{wave} of a pandemic?'', or ``Is there a \emph{vortex} in a given flow and what is its spatial extent?'', or questions about the presence of complex, multifaceted conditions, for example in medicine. With regards to the first question, although everyone has a clear idea of a `wave', different mathematical criteria are conceivable that capture the mental construct. Also, everyone has a vivid idea of what a flow vortex is; nevertheless, even decades of research have not led to a full agreement on what is the best criterion for determining the presence and extent of a flow vortex.
	
	%%%
	However, if data science (including data visualization) wants to answer such questions on base of data, precise and formalized definitions of such concepts are required.
	Especially in light of the fact that the increasing amount of data requires more and more automatic processing, which also calls for machine-understandable definitions of concepts.
	Precise mathematical definitions of concepts are also necessary for the creation of phenomenological models, i.e., mathematical models that contain such concepts as building blocks.
	
	%%%
	The question is: How can one build a bridge from the world of mental concepts, which are at best described verbally, to the world of precise, formal definitions?
	To do this, one first needs a deep and ideally also mathematically oriented insight into the problem domain.
	This often enables experts to suggest a first guess for a formal definition.
	But, given a suggestion for such a definition, how do you assess its quality and how do you improve it?
	Here, visualization helps: it allows the comparison of structures that result from data and experimentally formulated formal definitions with the mental image of that structure.
	This enables an iterative improvement of formal structure definitions, considering also the mental images of different people or even an entire expert community \citep{Hege_distillation_2011}.
	
	The problem can also be considered from the point of view of uncertainty: The lack of a precise definition results in a non-negligible ``uncertainty in definition''.
	The uncertainty arises because it is not clear a priori which structure indicators are the best, how the structure definition should be defined mathematically, and which parameter values in definitions are preferable.
	The definitional uncertainty can be visualized, e.g., by displaying ensembles of structures that result from different variants of structure definitions and different data sets.
	And it can be reduced by an iterative, visually supported process.
	
	%%%
	The long-term goal (which might take decades to achieve) is, of course, to find definitions that are as ``objective'' as possible and that work for as many data sets as possible.
	We give advice on which principles should be observed in this regard. An essential role plays the selection of suitable indicator quantities that show the presence of a structure.
	Another problem is that only in exceptional cases definitions can  be found that are parameter-free.
	So, many definitions contain threshold values. How should these be determined?
	
	In this paper, we describe a first attempt to a systematic approach for finding such definitions in a visually supported, interactive way. 
	Here, we focus on spatiotemporal structures. 
	The presented tool makes it possible to compare various definitions, indicator quantities, and parameter ranges and thus allows us to achieve suitable definitions in an iterative process. It can be used both for the quick, pragmatic finding of a suitable definition for given data sets, as well as for the long-term goal of finding a definition that is as universal as possible.
	We demonstrate the use of the tool by applying it to two meteorological examples: finding structure definitions for the vortex cores and centerlines, respectively, of temporarily evolving tropical cyclones.
	
	%%%
	Overall the paper presents a systematic approach and workflow as well as an interactive tool that enables expert users
	\begin{itemize}
		\item to find formal definitions of spatiotemporal structures, so that their properties correspond for all time steps as closely as possible to those of the mental image;
		\item to evaluate the suitability of different variables that indicate the presence of a structure;
		\item to narrow down the intervals of the parameters in a structure definition such that the structure properties match the mental image, and the variance of the resulting structures becomes as small as possible.
	\end{itemize}
	The proposed tool utilizes parallel coordinates with time as one variable to simultaneously depict parameter intervals and structure attributes for selected time intervals. In contrast to previous approaches for interactive structure extraction (e.g.\ using brushing \& linking), it directly supports the evaluation of all influencing factors, namely structure definitions (including computational methods for structure extraction), type of indicator quantities, and parameter intervals.
	
	%%%     
	The paper is organized as follows: In Sect.~\ref{sect:related-work}, we present related work.
	Sect.~\ref{sect:structure-definitions} focuses on describing the overall systematic approach as well as the prototype tool we implemented.
	In Sect.~\ref{sect:applications}, we apply our approach and tool to two use cases from meteorology.
	Finally, Sect.~\ref{sect:discussion} concludes the paper with a discussion of our findings and potential future work.

	%-------------------------------------------------------------------------
	
	\section{Related work}
	\label{sect:related-work}
	
	Concepts are the building blocks of thoughts and are necessary for the  representation of knowledge as well as for mental processes like categorization, inference, decision making, learning and communication.
	This view has become widely accepted in cognitive science and philosophy of science, see e.g., \citep{gardenfors2004conceptual,zenker2015applications}.
	Here we perceive concepts as mental constructs (as opposed to abstract objects or skills) and we use a naive notion of concepts based on concrete examples, thus avoiding complex philosophical questions (see, e.g., \citep{margolis2021sep-concepts} for an introduction on this).
	
	% structure identification and extraction
	The extraction of structures from data has a long history in data visualization.
	The structures are often mathematical features.
	Most prominent examples are mathematical features of spatial or spatiotemporal fields, like extremal structures, level sets, critical points and skeletons of stable manifolds.
	Therefore, the term ``feature extraction'' has become established in visualization.
	When dealing with application-oriented structures that carry semantics, such as jet streams, precipitation fields or tropical storms, it is better to speak of ``structures'' or ``concepts''. These are characterized by certain properties, which may include also mathematical features.
	This terminology corresponds to that of many other sciences, but differs from that traditionally used in visualization, where application-oriented structures are also often called ``features''.
	
	For a survey on the substantial history of structure extraction and visualization, see reviews on geometric-topological objects~\citep{mcloughlin2010geometric, pobitzer2011state}, on coherent set detection~\citep{hadjighasem2017critical}, on feature tracking in meteorological contexts~\citep{clark2014application,cai2015object}, and on the related problem of image segmentation~\citep{elbaz2017imagesegmentation}.
	Most approaches in data visualization refer to structures defined by single scalar or vector fields, e.g., level-sets or ridges of Lyaponov exponents or, more application-related, the centerlines of vortices or jet streams \citep{kern2018robust}.

	Data visualization often focuses on isolating application-defined regions of interest in spatiotemporal data. In a multidimensional ``attribute space'', defined by the value ranges of given and derived data, one searches interactively for subregions containing the value combinations of interest.
	Related procedures have been developed for image segmentation and volume rendering, where transfer functions map data characteristics to optical qualities, like color and opacity, to highlight regions of interest \citep{kniss2001interactive, ljung2016state}.
	
	An essential prerequisite for finding suitable structure definitions with visual control is interactivity. Here we can build on the research thread that began with the work by \citet{doleisch2003interactive}.
	In this, the definition of subregions in multivariate range space is greatly facilitated by multiple coordinated views (histograms, scatter plots, parallel coordinates, or function graphs), representing different variables side by side. When data points are interactively selected (``brushed'') in one view, the associated data items are immediately highlighted in all linked views.
	This ``linking'' is particularly powerful when attribute views are combined with 3D views, highlighting the corresponding preimages in the domain.
	This technique can also be extended to time-dependent data \citep{doleisch2004time-dependent, doleisch2006time-dependent} and has been applied to simulation data of Hurricane Isabel \citep{doleisch2004interactive-Isabel}.
	While in this work  all structures are defined by ranges of multivariate quantities, it is not concerned with finding more general and also parameter-dependent definitions of structures.
	
	A tool for interactive exploration of sets of parameter-dependent 3D geometries has been presented in \citep{beham2014cupid}. However, this work also did not aim at obtaining structural definitions.
	To the best of the authors' knowledge, there is no tool to date that supports the explicit goal of obtaining mathematical definitions of mentally and/or verbally given structures by using visual data analysis techniques.
		
	Visual comparison of similar, but different spatial data has been more deeply explored in the context of ensemble visualization, see, e.g., \citep{kolesar2017fractional} and references therein.
	
	The difficulty of finding a generally accepted, precise definition of concepts that is as objective as possible (i.e., independent of the person observing) is best illustrated by the example of flow vortices. Data visualization, together with flow research, has contributed a great deal to this. There is not enough space here, to sketch the nearly half-century-long research thread; instead we refer to the current state of research~\citep{theisel2021vortex} and the papers cited therein.
	In recent work~\citep{vonlindheim2021definition}, a formal approach to the definition and persistence of meteorological structures has been taken. This is exactly the kind of research we want to support and facilitate with the proposed procedure and interactive tool.

	%%%
	\section{Identifying structure definitions}
	\label{sect:structure-definitions}
	
	In the following, we first describe the proposed general procedure to identify structure definitions from mental images.
	Following this, we will present an interactive tool that we then apply to concrete examples in Sect.~\ref{sect:applications}.
	
	\subsection{General procedure}
	\label{sect:structure-definitions:general-proc}
	
\begin{figure}[tb]
		\centering
		\includegraphics[width=.8\linewidth]{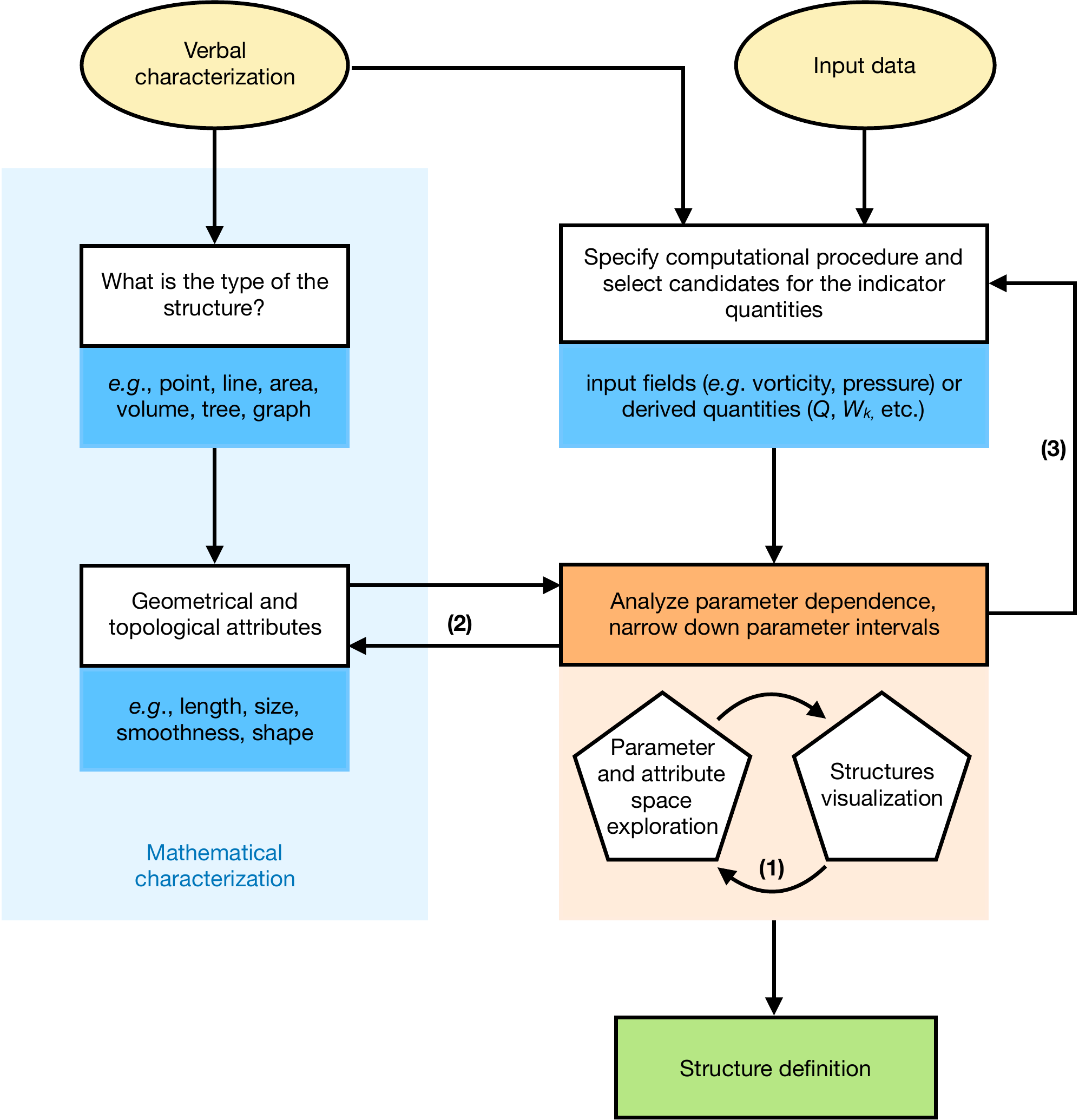}
		\caption{Schematic representation of proposed procedure.
			The numbers at the arrows refer to the three different types of iterative refinement that might be needed to arrive at a final structure definition; please see the text for details.
		}
		\vspace{-0.3cm}
		\label{fig:approach_scheme}
	\end{figure}

	The general structure of the proposed systematic approach is shown in Fig.~\ref{fig:approach_scheme}.
	It represents an iterative workflow that allows us to identify suitable structure definitions that could be further narrowed as more data comes in.
	
	The workflow starts with a verbal characterization, or a mental image, of the structure that we want to define more precisely in mathematical terms.	
	
	Based on the verbal characterization and our knowledge of mathematical structures, we can describe the structure using mathematical terms, e.g., the structure is a point, a line, an area, etc.
	The next step is then to identify attributes, in our case, geometric and topological attributes, that are well suited to describe the structure in a more or less complete and accurate manner.
	In particular, we are interested in attributes that allow us to distinguish structures matching our mental image from structures that do not.
	This part of the workflow is shown on the left side of Fig.~\ref{fig:approach_scheme}.
	
	The second input to our workflow is some kind of data, e.g., fields, that contain or describe the structure for which we want to find a definition.
	Sometimes, one does not work directly on the original input data, e.g., the velocity field, but on derived data, like $Q$ or $\Omega$ (Sect.~\ref{sect:applications}).
	For the sake of generality, let us call all data that indicate the presence of the structure in question, be it the original or derived data, indicator quantities.
	The set of such indicator quantities can be limited by using application-based higher-level principles, e.g., invariances or non-dimensionality.
	In order to extract the structure from one or more indicator quantities, computational procedures are needed.
	Usually, such a procedure depends on parameters that will influence the appearance of the extracted structure.
	An algorithmic structure definition thus consists of a computational procedure, one or more indicator quantities, and suitable computational parameters (or parameter ranges).
	
	In a final step, the algorithmic definition given by the computational procedure is translated back into a precise mathematical definition. For example, applying the Marching Cube algorithm to an indicator field is equivalent to applying a threshold operator to a scalar field defined by a formula. After this back translation, a mathematical structure definition results, consisting of mathematical expressions and, if these are parameterized, suitable parameter values or ranges.
	
	Given potential computational procedures as well as indicator quantities for the extraction of the structure of interest, the task of finding a structure definition reduces to sampling the whole space of possibilities including computational procedures, indicator quantities, and computational parameters, and comparing the outcomes with our mental image.
	Visually assessing all sampled structures would be very time-consuming.
	Therefore, we want to reduce the overall number of samples to the ones that are most likely to match our mental image.
	Here, the structure attributes come into play.
	They are computed for each of the sampled structures and can subsequently be used for filtering.
	
	This is done in the next step of our workflow, which analyzes the parameter dependence of the structure.
	The essential part here is the interactive selection of parameter ranges based on selection in the structure attribute space.
	As result, potential computational procedures, indicator quantities, and computational parameters are identified that lead to structures matching the selected structure attributes.
	Finally, the thus extracted structures are then visualized and visually assessed.
	Hence, our visual perception plays the final role of quality control.
	Since we have narrowed down the parameter space by brushing in the attribute space, visual assessment is reduced to a small set of structures.
	This visual assessment (also see Fig.~\ref{fig:approach_scheme}), however, might reveal the necessity that (1)~the parameter selection needs to be refined, (2)~other attributes that better describe our mental image are needed, or that (3)~other computational procedures and/or indicator quantities are required.
	If only a refinement of the parameter selection is necessary, after some iterations we might arrive at a structure definition.
	Otherwise, new attribute descriptors and/or computational procedures will have to be implemented.
	
	In the next section, we will describe the interactive tool we use in our application to explore the attribute and parameter spaces.	

        %%%
	\subsection{Interactive tool}
	\label{subsec:pc_plot}
	
	\begin{figure*}[tb]
		\centering
		\includegraphics[width=1\linewidth]{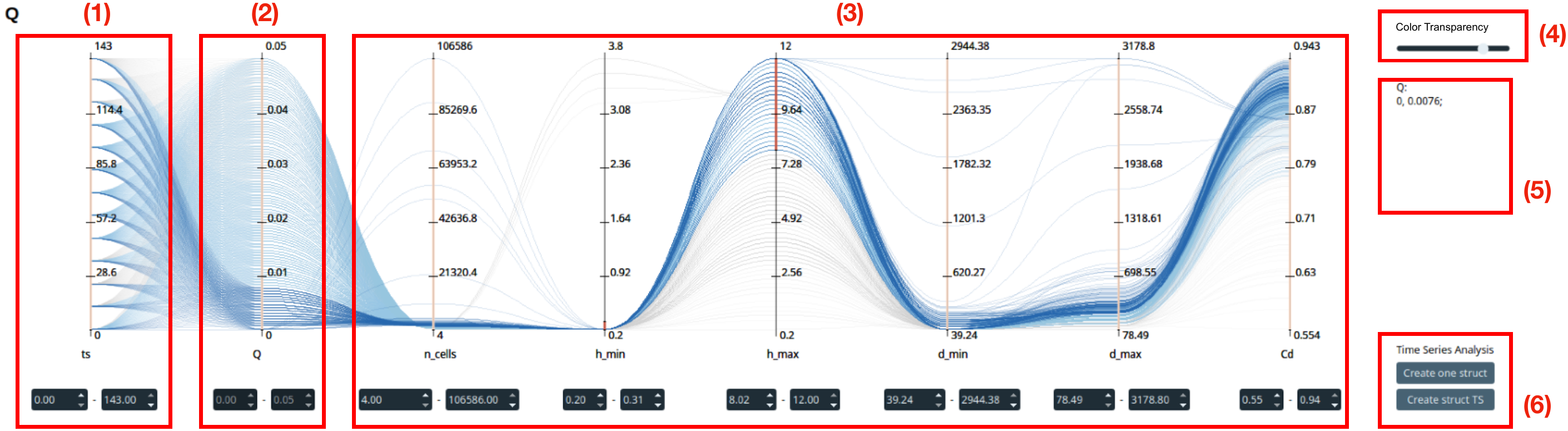}
		\caption{Parallel coordinates window and its main areas: (1)~time axis; (2)~parameter space; (3)~attribute space; (4)~color/transparency settings; (5)~quantitative representation of selected parameter intervals; and (6)~buttons for interactive exploration of selected structures.
		}
		\vspace{-0.3cm}
		\label{fig:pc_plot_structure}
	\end{figure*}

	We utilize the parallel coordinates (PC) visualization for parameter and attribute space exploration.
	This technique allows interactive investigation of multidimensional data.
	It has been successfully used in many application.
	Most closely related is probably the work by~\citet{beham2014cupid} for exploring the parameter space of a geometry generator.
	
	The PC plot is shown in Fig.~\ref{fig:pc_plot_structure}.
	A single line in the plot corresponds to one structure sampled from the parameter space of one time step, accompanied with a set of computed attributes.
	Hence, in our application, the high-dimensional space that is visualized using parallel coordinates consists of the time dimension, the dimensions of the parameter space, and the dimensions of the attribute space.
	The plot's first axis represents time (Fig.~\ref{fig:pc_plot_structure}~(1)).
	Brushing on the time axis could be done to restrict the considered time range to a subrange.
	However, since in our applications we are interested in parameters that are suitable for the whole time range, we do not use this functionality.
	The second axis represents the parameter space~(Fig.~\ref{fig:pc_plot_structure}~(2)).
	Note that we could potentially have more than one parameter axis depending on the dimensionality of the parameter space.
	The parameter axes display only the result of interactive brushing in the attribute space~(Fig.~\ref{fig:pc_plot_structure}~(3)).
	Brushing is done on each attribute axis separately.
	The selections from the individual brushing on all the attribute axes are concatenated to obtain one final selection. 
	Note that for the concatenation of the selection, currently only the logical AND operation is supported.
	As result of the brushing, the lines corresponding to the selected attribute values are colored, by default in a light blue.
	Transparency and color of the lines can be adapted~(Fig.~\ref{fig:pc_plot_structure}~(4)).
	The selected lines are analyzed w.r.t.\ their parameter values and parameter intervals are identified that are valid over the whole selected time range.
	These ranges are shown on the right sidebar~(Fig.~\ref{fig:pc_plot_structure}~(5)) and the corresponding lines are highlighted in dark blue to distinguish them from the rest of the selected lines.
	Lines that are not selected are shown in gray.
	The PC plot is connected with a direct visualization that allows simultaneous investigation of the selected structures~(Fig.~\ref{fig:pc_plot_structure}~(6)).
	
	%-------------------------------------------------------------------------
	\section{Applications}
	\label{sect:applications}
	
\begin{figure}[tb]
	\centering
	\includegraphics[width=0.9\linewidth]{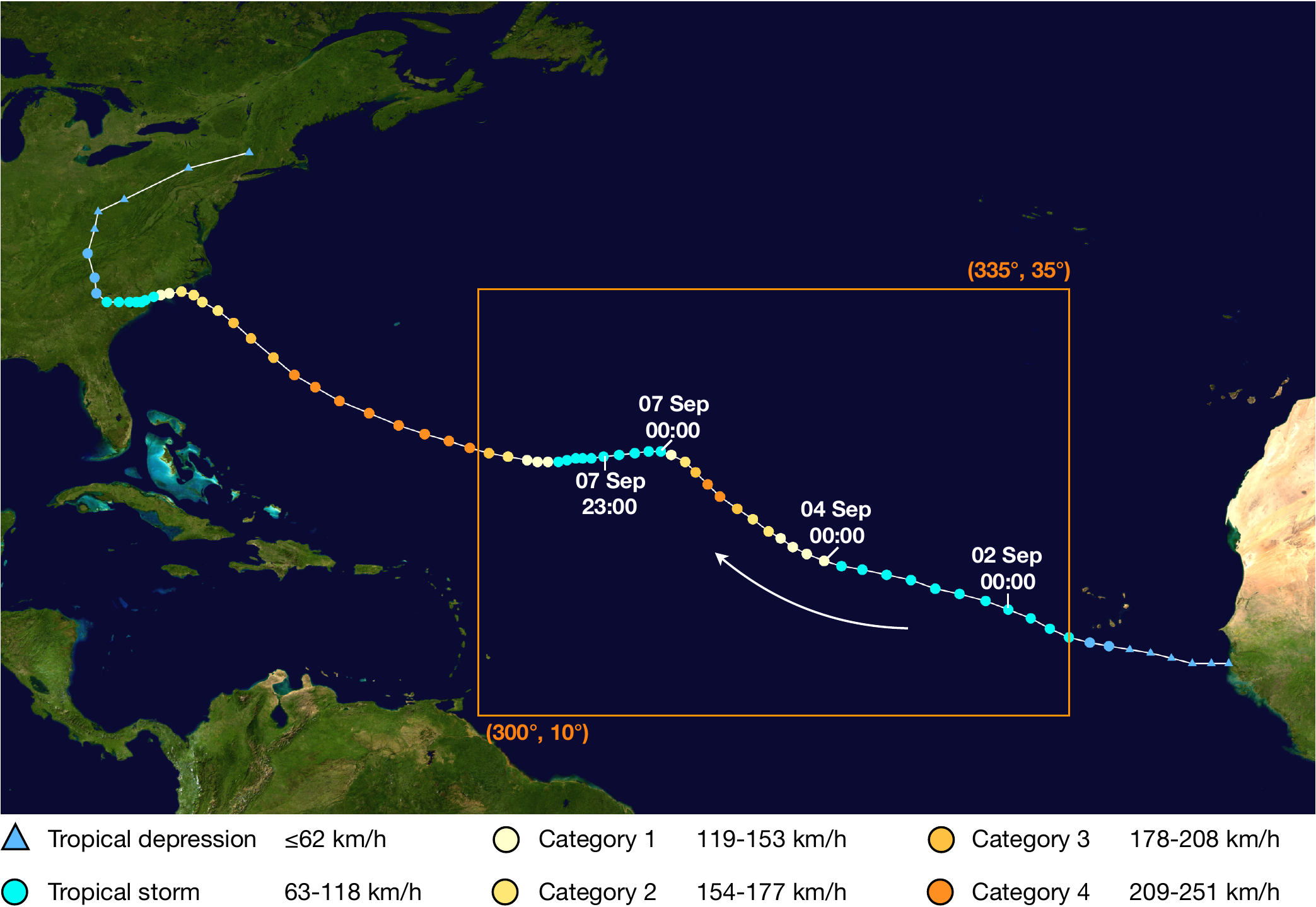}
	\caption{Track map of Hurricane Florence 2018 (basemap and track taken from \citep{wiki:Florence2018}). The points show the location of the storm at 6-hour intervals. The color represents the storm's maximum sustained wind speeds according to the Saffir–Simpson scale (see legend). The orange box defines the area of interest. }
	\vspace{-0.3cm}
	\label{fig:florence_track}
\end{figure}
	
	In this section, we give a clear and concise example of our approach by demonstrating the procedure for specific structures, namely spatio-temporal geometric structures of tropical cyclones (TC), i.e., their vortex cores and centerlines.
	The example is motivated by a new theory of TC intensification, according to which the inclination of the centerline, in interaction with atmospheric heat patterns that adopt some spatial orientation to the inclined centerlines, is a key parameter for TC intensification~\citep{paschke2012motion,doerffel2020stabilization,dorffel2021dynamics}.
	
	The analysis is made on basis of ERA5 reanalysis data of the Hurricane Florence 2018.
	The data is given on a regular horizontal grid with resolution of $\ang{0.25} \times\ang{0.25}$, i.e., about \qty{30}{\kilo\meter}~\citep{era5pres,era5single}. In all our visualizations, the data are vertically scaled with a factor of 100 to improve the perception of the structures.
	The time frame for the analysis is chosen to cover three different phases of the TC evolution: (1)~stages of intensification; (2)~full maturity; (3)~weakening.
	The track of Hurricane Florence and the region of interest are shown in Fig.~\ref{fig:florence_track}.

	\subsection{TC vortex core regions}
	
	The  word  ``vortex'' is associated with a whirling or circular motion.
	But it is still the case that ``no single definition of a vortex is currently universally accepted, despite the fact that fluid dynamicists continue to think in terms of vortices''~\citep{chakraborty2005relationships}.
	The same, in fact, is true for the question of ``what a TC is''.
	There are only vague notions about the flow being circular, the presence of a depression, or the cloud structure~\citep{Houze2009}.
	Vortices can be characterized by high local vorticity, more rigid-body rotation than stretching or shearing, low local pressure, closed or spiraling streamlines or pathlines, and coherent motion of neighboring fluid particles~\citep{epps2017review}.
	
	\subsubsection{Indicator quantities for vortex core extraction}
	\label{subsect:vortex-methods}
	
\begin{figure}[tb]
		\centering
		\includegraphics[width=0.7\linewidth]{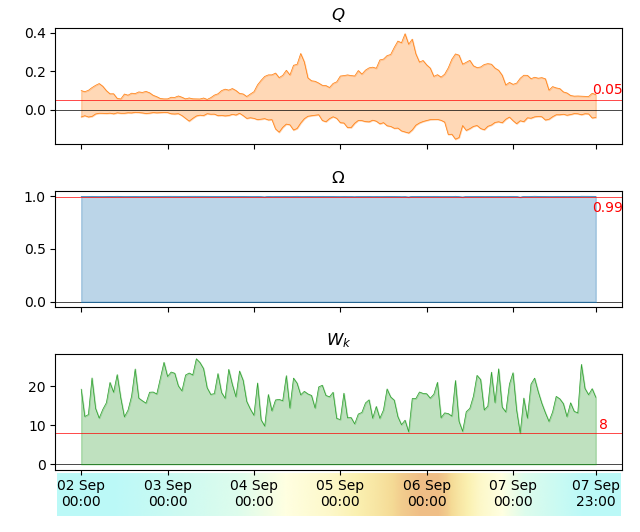}
		\caption{ Minimum and maximum values for three chosen indicator quantities in the time frame of 02-07 September 2018.
			Red line represents the upper limit of analysis intervals.
			Color gradient in the axis area represents the intensity of TC (see legend in Fig.~\ref{fig:florence_track}).}
		\vspace{-0.3cm}
		\label{fig:indicator_intervals}
	\end{figure}

	There is an abundance of methods for the extraction of vortex cores.
	Even though many of them are based on decomposing the velocity gradient $\nabla u$ into  strain-rate tensor $A=\frac{1}{2}(\nabla u+\nabla u^\intercal)$ and vorticity tensor $B=\frac{1}{2}(\nabla u-\nabla u^\intercal)$, they may provide different results.
	\begin{itemize}
		\item The widely used $Q$-criterion~\citep{hunt1988eddies} identifies a vortex as a ``connected fluid region with a positive second invariant of $\nabla u$''~\citep{kolar2007vortex} , i.e.,
		\begin{equation}
		\label{ineq_Q}
		Q = \dfrac{1}{2}(\norm{B}^2-\norm{A}^2)>0.
		\end{equation}
		\item $\Omega$ was introduced to represent the ratio of vortical deformation over the whole deformation inside a vortex core~\citep{liu2016omega}, i.e.,
		\[\Omega = \dfrac{\norm{B}^2}{\norm{B}^2+\norm{A}^2}.\]
		To avoid division by 0, it was proposed~\citep{liu2018epsilon} to add a small positive number $\epsilon_{\Omega}=0.001(\norm{B}^2-\norm{A}^2)_{\max}$ to the denominator such that the criterion becomes
		\begin{equation}
		\label{ineq_Omega}
		\Omega = \dfrac{\norm{B}^2}{\norm{B}^2+\norm{A}^2+\epsilon_{\Omega}}>0.5.
		\end{equation}
		The authors claim that this method ``is pretty universal and does not need much adjustment in different cases and the isosurfaces of $\Omega=0.52$ can always capture the vortices properly''~\citep{liu2016omega}.
		However, in order to capture the vortex core region, further adjustment of the threshold is needed.
		$\Omega$ is both dimensionless and normalized, resulting in values that are restricted to the interval $[0,1]$.
		\item The kinematic vorticity number~\citep{truesdell1953two}
		$W_k = {\norm{B}}/{\norm{A}}$
		is another promising tool for the determination of vortex properties~\citep{schielicke2016kinematic}.
		If the rotation rate prevails over the strain rate, $W_k$ is larger than 1.
		Early circulations that have a potential to develop into stronger vortices could be captured with a lower threshold of $W_k$ (e.g., $W_k$ slightly smaller than 1).
		On the other hand, a higher threshold of $W_k$ focuses on already developed strong vortices.
		To avoid non-physical noise, in our calculation we use an $\epsilon_{W_k} = 0.03(\norm{B}-\norm{A})$.
		The vortex core is then identified by
		\begin{equation}
		\label{ineq_Wk}
		W_k = \dfrac{\norm{B}}{\norm{A}+\epsilon_{W_k}} > 1.
		\end{equation}
		Similar to $\Omega$, $W_k$ is dimensionless, but it is not normalized.
		Hence, values can potentially fall into the interval $[0,\infty]$.
	\end{itemize}
	All three indicator quantities describe vortices, but their application may result in different vortex regions.
	\begin{figure}[tb]
		\centering
		\includegraphics[width=0.49\linewidth]{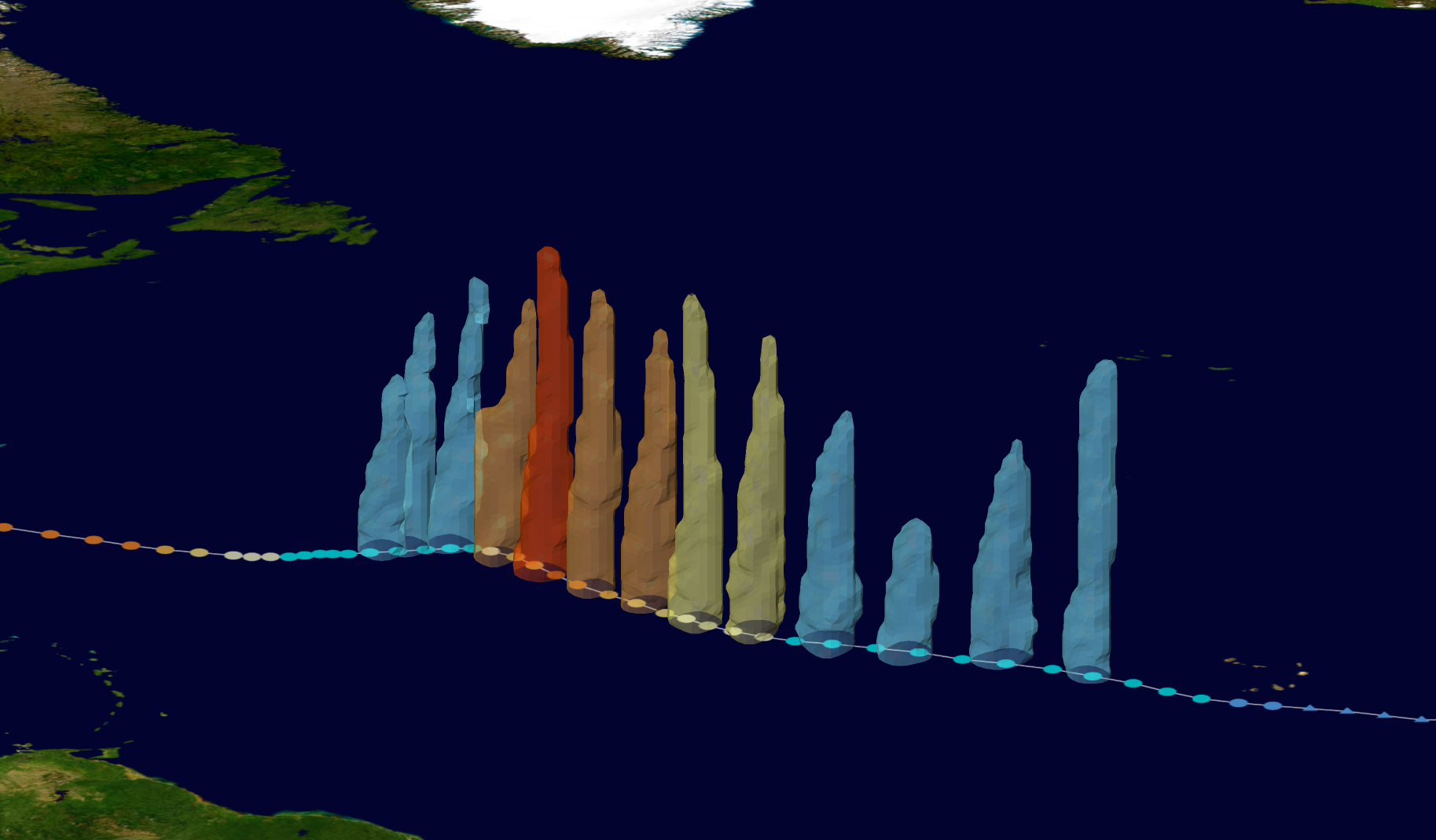}
		\hfill
		\includegraphics[width=0.49\linewidth]{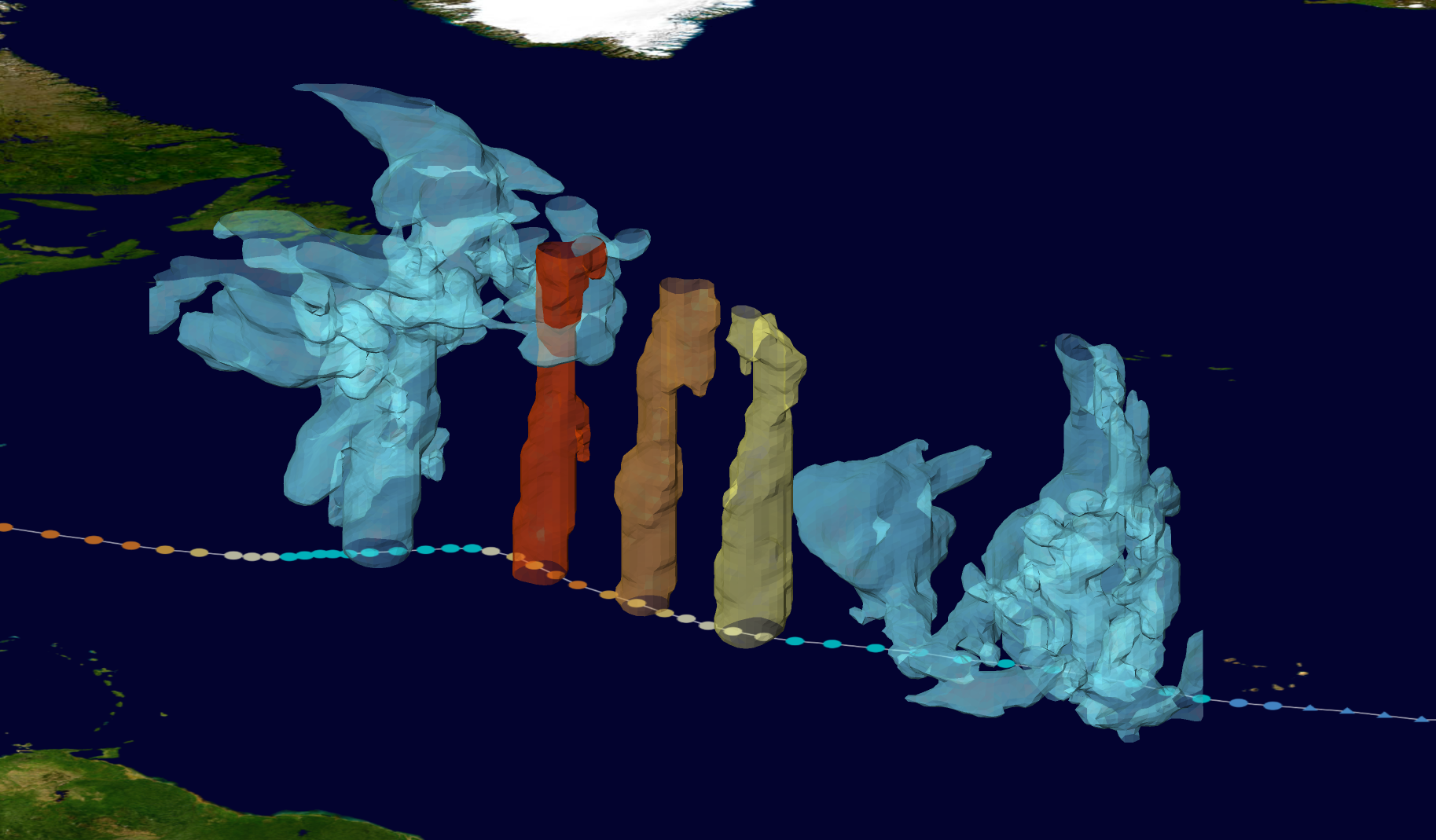}
		\caption{Vortex regions extracted using $Q>0.02$ (left) and $\Omega>0.52$ (right). The top image shows extracted vortex regions for every 12 hours from the time frame of 02-07 September.
			For $\Omega$, the following time steps were chosen: 02 September 00:00, 04 September 00:00, 05 September 00:00, 06 September 00:00, 07 September 23:00.
			Colors represent the intensity of TC according to the Saffir-Simpson scale (see legend in Fig.~\ref{fig:florence_track}).}
		\vspace{-0.3cm}
		\label{fig:motivation}
	\end{figure}
	The inequalities in Eq.~\ref{ineq_Q}-\ref{ineq_Wk} provide only theoretical thresholds.
	For real-case analyses, these values need to be adjusted to capture the appropriate structures.
	Fig.~\ref{fig:motivation} shows vortex regions extracted using $Q>0.02$ and $\Omega > 0.52$.
	Although at the highest hurricane intensity (yellow and orange structures), both quantities can capture the structures of the vortex core well, there are some issues during the intensification and weakening phases.
	Structures extracted with $Q>0.02$ shrink up to \qty{4.6}{\kilo\meter} while regions defined with $\Omega > 0.52$ do not represent TC cores correctly.
	In  Fig.~\ref{fig:indicator_intervals}, the value intervals for three indicator quantities over the time frame of interest are shown~(Fig.~\ref{fig:florence_track}).
	$\Omega$ lies always in $[0,1]$ while the extremum values for $Q$ and $W_k$ change over time.
	For the further analysis, parameter values that do not exist are excluded.
	The upper limits of the selected parameter intervals are depicted by a red line in Fig.~\ref{fig:indicator_intervals}.
	
	\subsubsection{Geometrical attributes of vortex regions}
	\label{sssect:prop_region}
	
	A vortex core region is associated with a dense cylinder-like structure.
	Based on this mental image and the examples shown above, the following attributes were chosen to describe the geometry of the extracted vortex core regions:
	\begin{center}
		\begin{tabular}{ r l } 
			\hline
			$n_{cells}$ & number of voxels  \\ 
			$h_{min}$ & minimum height  \\ 
			$h_{max}$ & maximum height \\ 
			$d_{min}$ &  minimum diameter \\ 
			$d_{max}$ &  maximum diameter  \\ 
			$C_d$ & compactness value \citep{bribiesca2008compactness} \\ 
			\hline
		\end{tabular}
	\end{center}
	The \textit{number of voxels} describes the overall size of the structure.
	\textit{Minimum} and \textit{maximum heights} give the lowest and the highest levels where the structure is defined.
	These attributes help to ensure that the extracted structure starts at a very low level and reaches a large height.
	The diameters of the structure were computed for each individual horizontal slice using the formula
	\[
	d = \sqrt{dX_{max}^2+dY_{max}^2},
	\]
	where $dX_{max}$ and $dY_{max}$ are the maximum distances the structure spreads in $x$ and $y$ directions, respectively.
	The derived value $d$ approximates the diameter of a circle enclosing all selected voxels in the slice.
	It can be interpreted as the radius of the circumscribed circle.
	The \textit{diameter} attributes are derived as minimum and maximum values of $d$ over all horizontal slices. 
	\textit{Diameter} and \textit{compactness value} control the width of the structure and its shape. 
	The latter was introduced by~\citet{bribiesca2008compactness} and is defined as 
	\[ C_d = \dfrac{A_c}{A_{c_{max}}},\]
	where $A_c$ is the contact surface area and $A_{c_{max}}$ the maximum contact surface area.
	For a solid composed of $n$ voxels, the approximation $A_{c_{max}} \approx 3(n-(n)^\frac{2}{3}) $ can be used.
	The measure of discrete compactness is dimensionless, has values from 0 to 1 and is maximized by a cube.
		
	\begin{figure}[tb]
		\centering
		\includegraphics[width=0.6\linewidth]{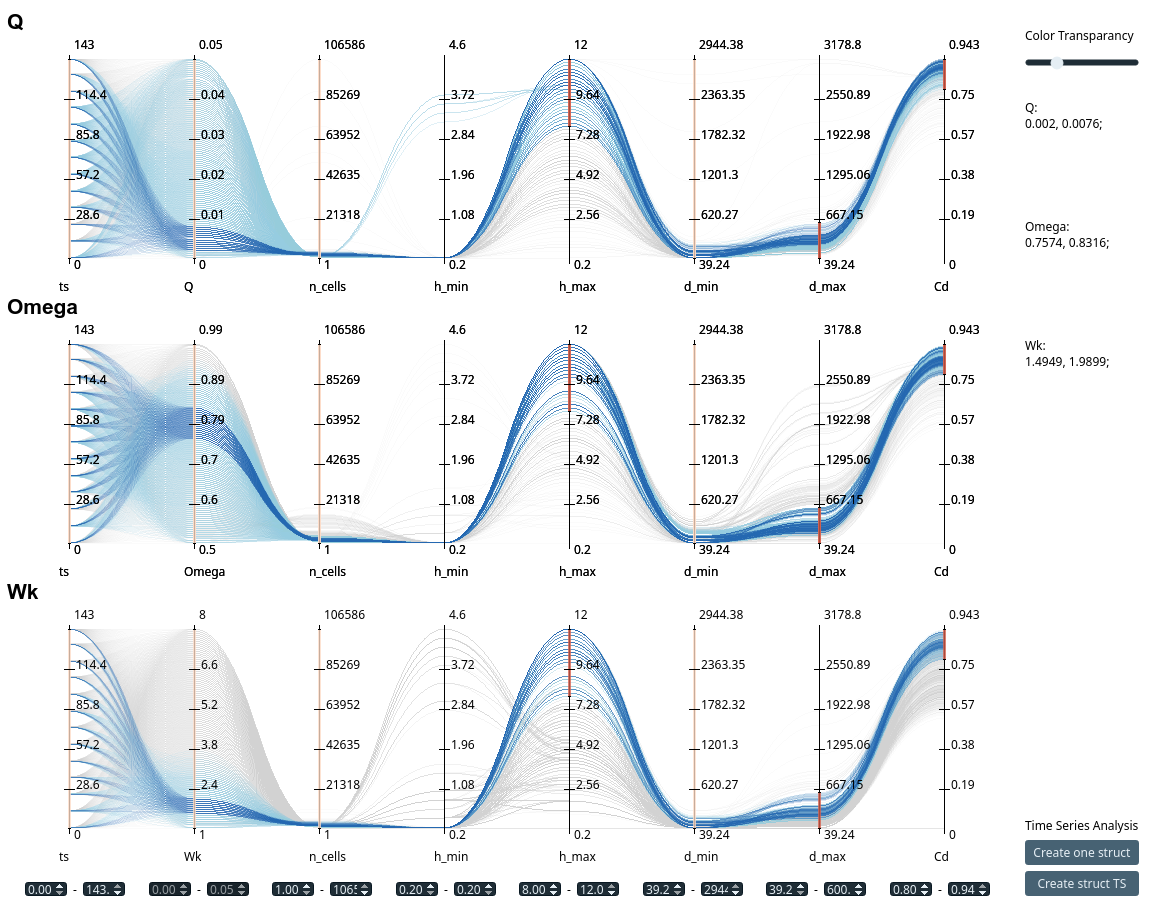}\\
		\includegraphics[width=0.6\linewidth]{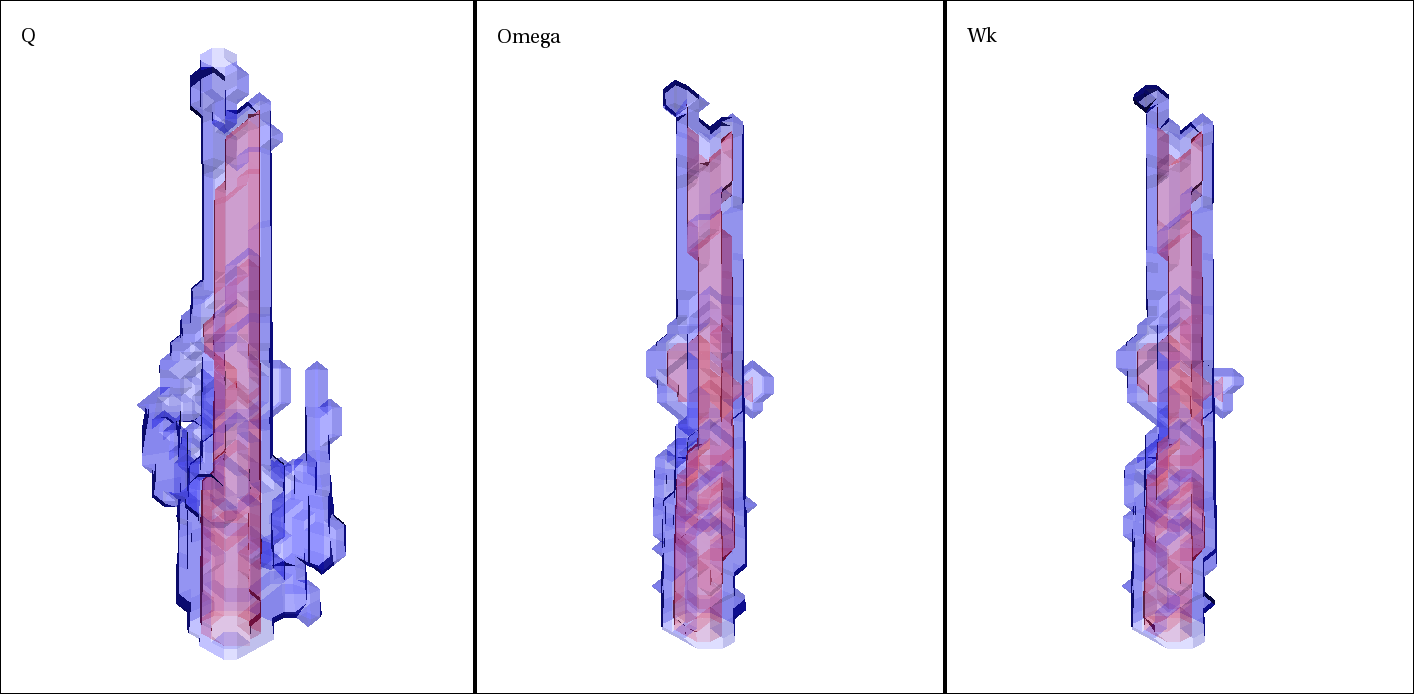}
		\caption{\label{fig:pc_3indicators}
			Parallel coordinates plot window and structures extracted by selecting proper parameter values for three indicator quantities $Q$, $\Omega$ and  $W_k$. Red and blue isosurfaces correspond to smallest and biggest structures respectively. }
		\vspace{-0.0cm}
	\end{figure}
	
	\subsubsection{Analysis and results}
	
	Three indicator quantities from Sect.~\ref{subsect:vortex-methods} were computed based on the velocity field $u$ for 13 time steps in time interval 02-07 September 2018.
	For each indicator quantity and each time step, sample structures were derived using 100 evenly distributed values from the appropriate parameter interval, i.e.,
	\begin{itemize}  
	    \setlength\itemsep{0em}
		\item $Q \in [0,0.05]$
		\item $\Omega \in [0.5, 0.99]$
		\item $W_k \in [1,8]$
	\end{itemize}
	Properties describing each structure were computed as discussed in Sect.~\ref{sssect:prop_region}.
	To inspect the 3D structures defined by properties selected via the PC plot, isosurfaces of the minimum and maximum structures are used, i.e., structures corresponding to the upper and lower limit of the derived parameter interval, respectively.
	
	PC plots with the desired attributes for the three indicator quantities are shown in Fig.~\ref{fig:pc_3indicators}.
	The vortex core should start at the first presented level \qty{0.2}{\kilo\meter} and have at least \qty{8}{\kilo\meter} height.
	To limit the horizontal spread, structures with maximum diameter exceeding \qty{600}{\kilo\meter} and with small compactness value, i.e., $C_d<0.8$, were excluded.
	Lines representing structures meeting all the constraints are shown in the PC plot by two shades of blue.
	Dark blue highlights the interval that would lead to the derivation of structures with the desired properties for every time step included in the analysis.
	
	The procedure leads to the following structure definitions for TC cores based on the indicator quantities $Q$, $\Omega$ and $W_k$:
	\begin{alignat*}{4}
		C_Q &= \{x \in \mathbb{R}^3\:|\: Q(x) &&>\tau_Q;&&\tau_Q&&\in [0.002,0.0076]\}, \\
		C_\Omega &= \{x \in \mathbb{R}^3\:|\: \Omega(x)&&>\tau_\Omega;&&\tau_\Omega&&\in [0.7574,0.8316]\}, \\
		C_{W_k} &= \{x \in \mathbb{R}^3\:|\:W_k(x)&&>\tau_{W_k};\:&&\tau_{W_k}&&\in [1.4949,1.9899]\}.
	\end{alignat*}
	
	The obtained definitions correspond to the ERA5 reanalysis data of Hurricane Florence 2018. Further analyses are necessary in order to get structure definitions that are more generally applicable.
	
\begin{figure}[tb]
		\centering
		\includegraphics[width=0.49\linewidth]{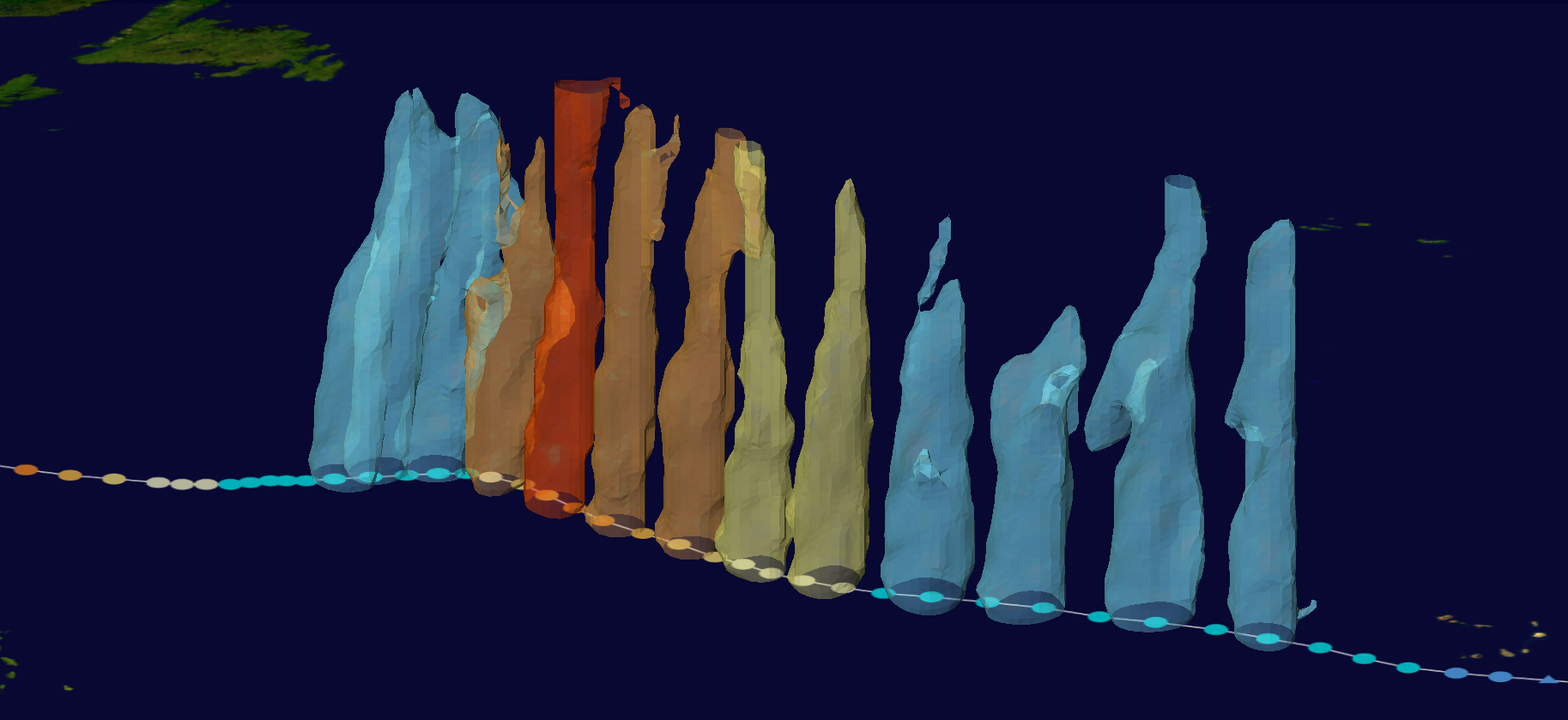}
		\hfill
		\includegraphics[width=0.49\linewidth]{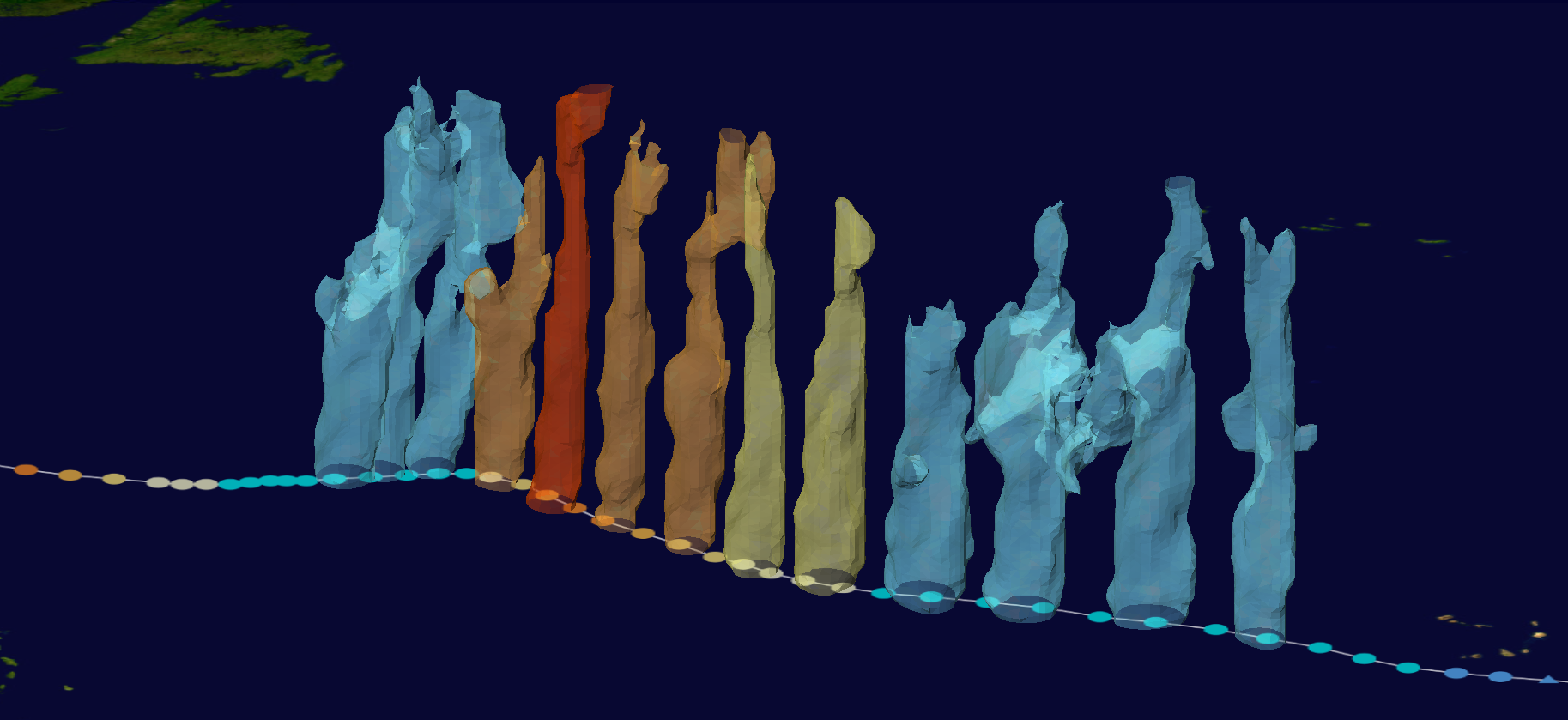}
		\caption{Vortex regions showing the average 
			structure from the derived intervals for $Q$ (left) and $\Omega$ (right).
			Time steps at every 12 hours of the time frame of 02-07 September are shown.
			Colors represent the intensity of TC according to the Saffir-Simpson scale (see legend in Fig.~\ref{fig:florence_track}).}
		\vspace{-0.0cm}
		\label{fig:result_iso}
	\end{figure}

	Some resulting structures found for $Q$ and $\Omega$ for the whole time window are shown in Fig.~\ref{fig:result_iso}.
	Both indicator quantities provide comparable vortex core regions.
	
	\subsection{TC core line}
	
	A TC core line is a structure that describes position and course of a TC.
	It is located in or close to the center of the TC vortex and represents another mental concept that is not explicitly described by data fields.
	Representation of the TC domain in cylindrical coordinates that are centered at the vortex core line is very useful for studying TC structure and dynamics (see \citep{paschke2012motion,dorffel2021dynamics}). Quantitative results, however, highly depend on a TC core line. 
	
	\subsubsection{Core line computation}
	
	Here, we extract TC core lines based on the TC vortex core regions that were discussed previously.
	In addition to the indicator quantities, the mathematical definition of the center should be considered.
	We focus on two methods for core line computation.
	Both methods compute one point per horizontal slice.
	
	\noindent\textbf{Extremum method:} This method identifies the center of a vortex as the extremum of an indicator quantity~\citep{sahner2005galilean}.
	Usually, the maximum is used; in some cases, e.g., for pressure fields, the minimum value may be used.
	
	\noindent\textbf{Centroid method:} This method depends on a search radius~$R$~\citep{Nguen2014Centerline}.
	The location of the minimum pressure on a constant height surface was used as first guess, and the centroid was calculated within a circle of radius $R$ around the first guess:
	\begin{equation}
	\label{eq:centroid}
	\Bar{x} = \frac{\sum\limits_{r<R}x_i\omega_i}{\sum\limits_{r<R}\omega_i} \text{ and } \Bar{y} = \dfrac{\sum\limits_{r<R}y_i\omega_i}{\sum\limits_{r<R}\omega_i},
	\end{equation}
	where $x_i$, $y_i$ are grid points within a circle of radius $R$ and $\omega_i$ is the values of indicator quantity at this point.
	In order to determine the center based on the quantity that should be minimized, inverted values could be used, e.g., pressure deficit $p' = \bar{p} - p$.
	Potential vorticity (PV) was proposed~\citep{Nguen2014Centerline} as indicator $\omega$ considering only points within the circle with positive PV.
	We combine this with the extracted structures, i.e., instead of determining the first guess center and using various radii $R$, we apply Eq.~\ref{eq:centroid} on the TC vortex core regions extracted using indicator quantities $Q$, $\Omega$, and $W_k$.

	\subsubsection{Geometrical attributes for a line}
	\label{sssect:prop_line}
	
	With the same motivation as in Sect.~\ref{sssect:prop_region}, we determined the following attributes for line-like features to help identify structures matching our mental image of a core line:
	\begin{center}
		\begin{tabular}{ r l } 
			\hline
			$h_{min}$ & minimum height  \\ 
			$h_{max}$ & maximum height \\ 
			$dX$ & overall horizontal displacement \\ 
			$\overline{dX}$ & mean of horizontal displacement \\ 
			$\max(dX)$ & maximum horizontal displacement \\ 
			$c$ & line curvature \citep{Leger1999MengerCA} \\ 
			\hline
		\end{tabular}
	\end{center}
	\textit{Minimum} and \textit{maximum height} are the smallest and largest $z$-coordinates over all line points.
	It corresponds to $h_{min}$ and $h_{max}$ defined for the TC vortex core region.
	Further properties, such as \textit{horizontal displacement} and \textit{line curvature}, are describe the smoothness of a line.
	\textit{Horizontal displacement} is computed as Euclidean distance between two 2D points $X_{z_i}$ and $X_{z_j}$, i.e.,
	\[dX_{ij}= d(X_{z_i}, X_{z_j}), \]
	where $z_i$ and $z_j$ are the corresponding horizontal levels.
	\textit{Overall horizontal displacement} is measured between the start and end points of the line.
	\textit{Mean} and \textit{maximum horizontal displacement} are the average and maximum values of $dX$, respectively, calculated for every two consecutive levels.
	Finally, \textit{line curvature} is defined as
	\begin{align*}
		c(A,B,C) &= 4\dfrac{\text{Area of the triangle}(A,B,C)}{d(A,B)d(B,C)d(C,A)}\\ 
		&= 2\dfrac{sin(\angle ABC)}{d(C,A)},
	\end{align*}
	where $A$, $B$ and $C$ are points that represents two consecutive line sections and $d(\cdot, \cdot)$ is the Euclidean distance in $\mathbb{R}^3$~\citep{Leger1999MengerCA}.
	
	\begin{figure}[tb]
		\centering
		\includegraphics[width=0.15\linewidth]{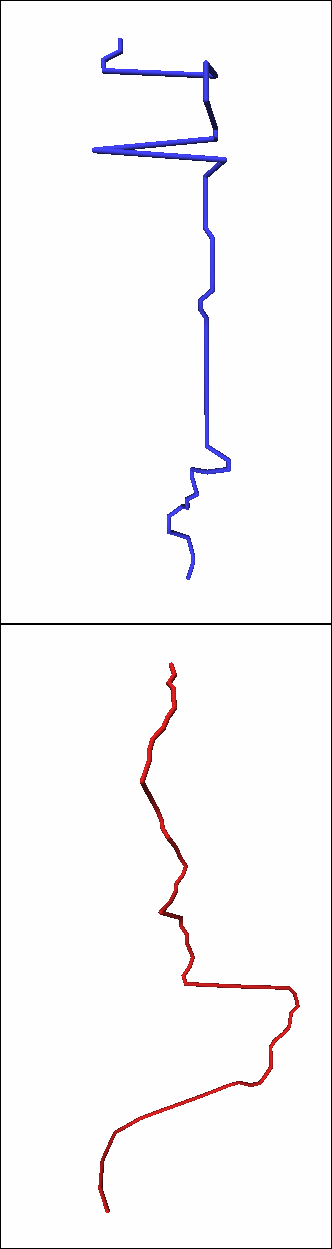}
		\includegraphics[width=0.15\linewidth]{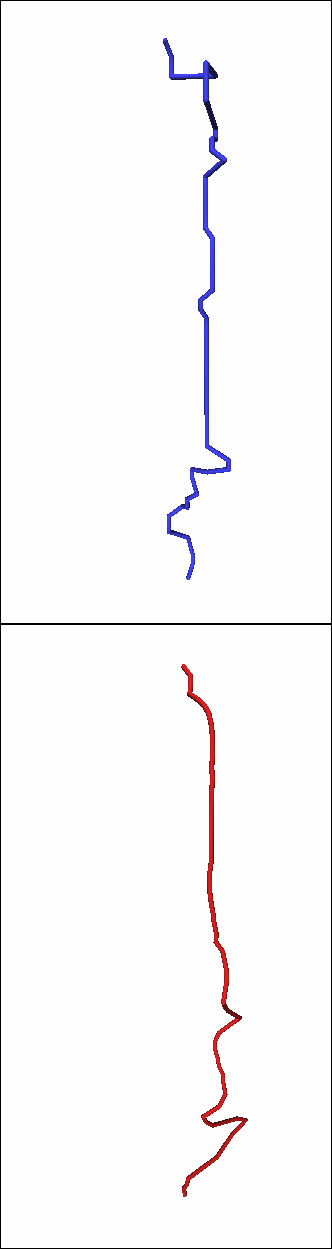}
		\includegraphics[width=0.15\linewidth]{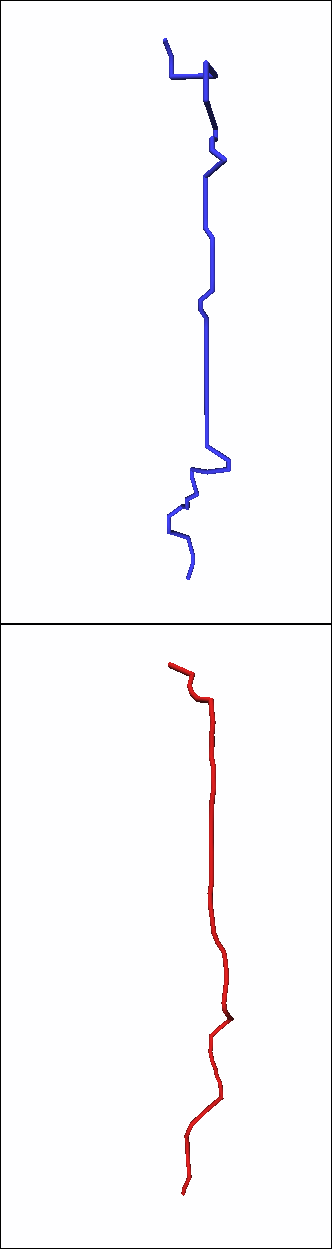}
		\includegraphics[width=0.15\linewidth]{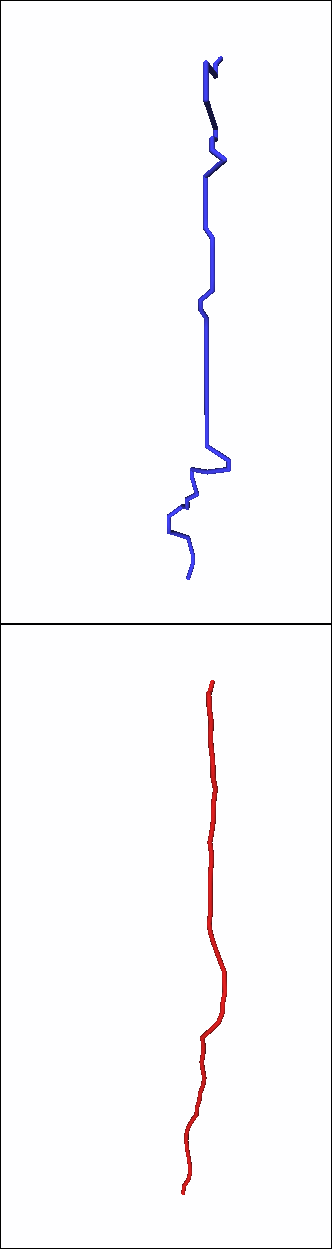}
		\caption{\label{fig:centroid_vs_max_lines}
			Lines extracted by maximum (blue lines, top row) and centroid method (red lines, bottom row) using - from left to right - $\Omega > 0.5$, $\Omega > 0.65$, $\Omega > 0.7$ and $\Omega > 0.8$.}
	\end{figure}

	\subsubsection{Analysis and results}
	
\begin{figure}[tb]
		\centering
		\includegraphics[width=0.6\linewidth]{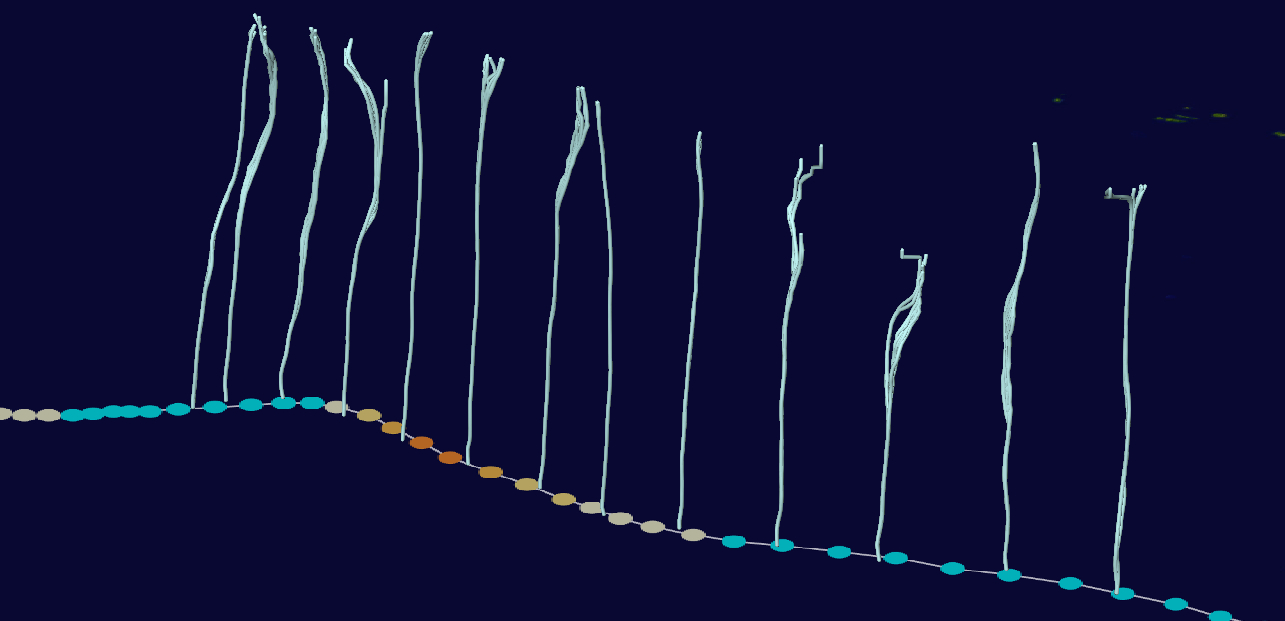}\\
		\includegraphics[width=0.6\linewidth]{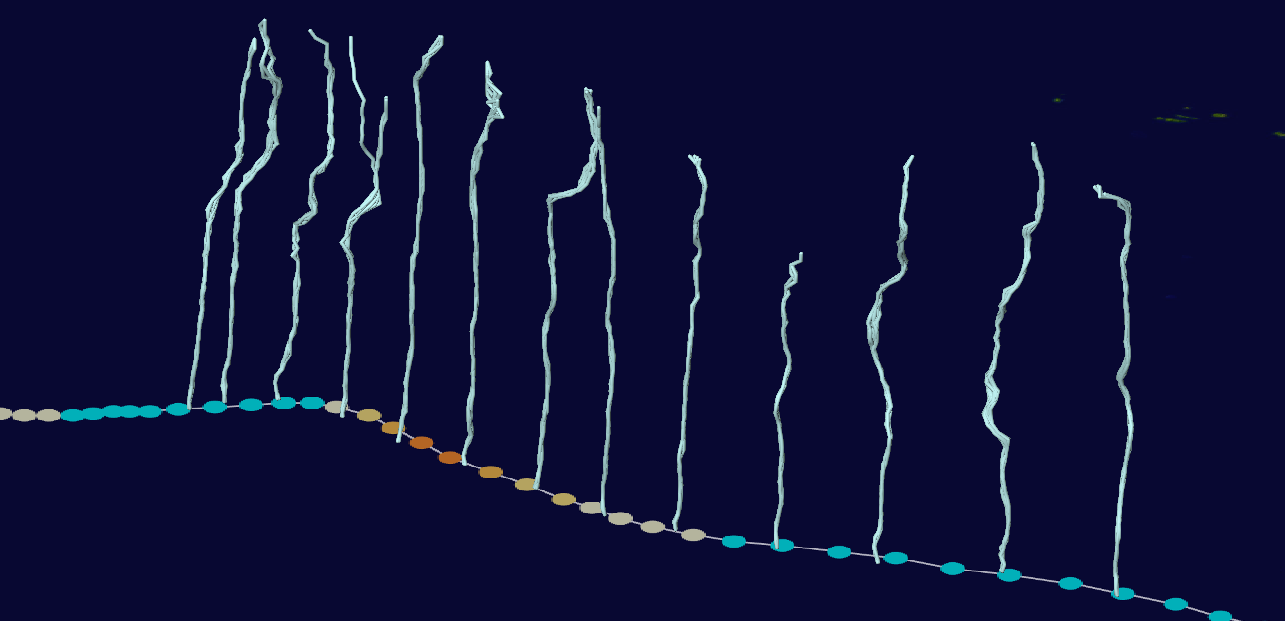}\\
		\includegraphics[width=0.6\linewidth]{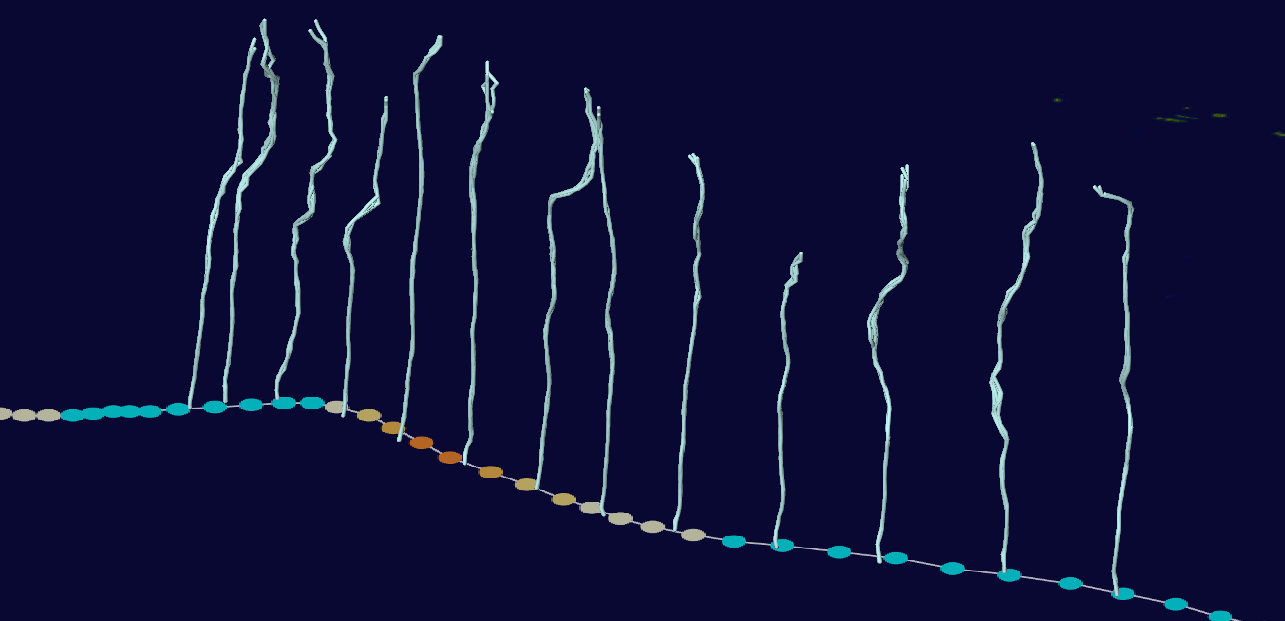}
		\caption{\label{fig:lines_ts}
			Core lines detected using the centroid method for $Q$ (top), $\Omega$ (middle), and $W_k$ (bottom), showing 9, 11, 5 lines, respectively.
			Selected time frame: 02-07 September.}
	\end{figure}

	First, we compare the two line extraction methods, that is, the extremum with the centroid method.
	Fig.~\ref{fig:centroid_vs_max_lines} shows some examples of the extracted structures using indicator $\Omega$.	
	The lines extracted using the extremum method show less variation; however, most of them are of zigzag form.
	At the same time, the centroid method provides smoother lines the shape of which can drastically change along with the parameters, as can be seen in Fig.~\ref{fig:centroid_vs_max_lines}, bottom row, left two images.
	The visualization of this example suggests that the structures extracted with the extremum method are worse than those extracted with the centroid method.
	Incorporating parameter space exploration with structure visualization can help objectify this observations.
	Interestingly, no common interval can be found for the extremum method for the same parameters that derive reasonable TC centers as with using the centroid method, see Fig.~\ref{fig:centroid_vs_max_lines}.
	Lines extracted for this parameter setting can be found in Fig.~\ref{fig:lines_ts}.
	To depict multiple lines at ones, we used the easiest but at the same time most informative technique where we show the lines together without further adjustments.
	The main issue with this visualization occurs if lines perfectly overlap.
	Therefore, this approach leads to an indistinct representation of structures found by the extremum method, while it worked well for the lines defined by the centroid method.
	\begin{figure}[tb]
		\centering
		\includegraphics[width=1\linewidth]{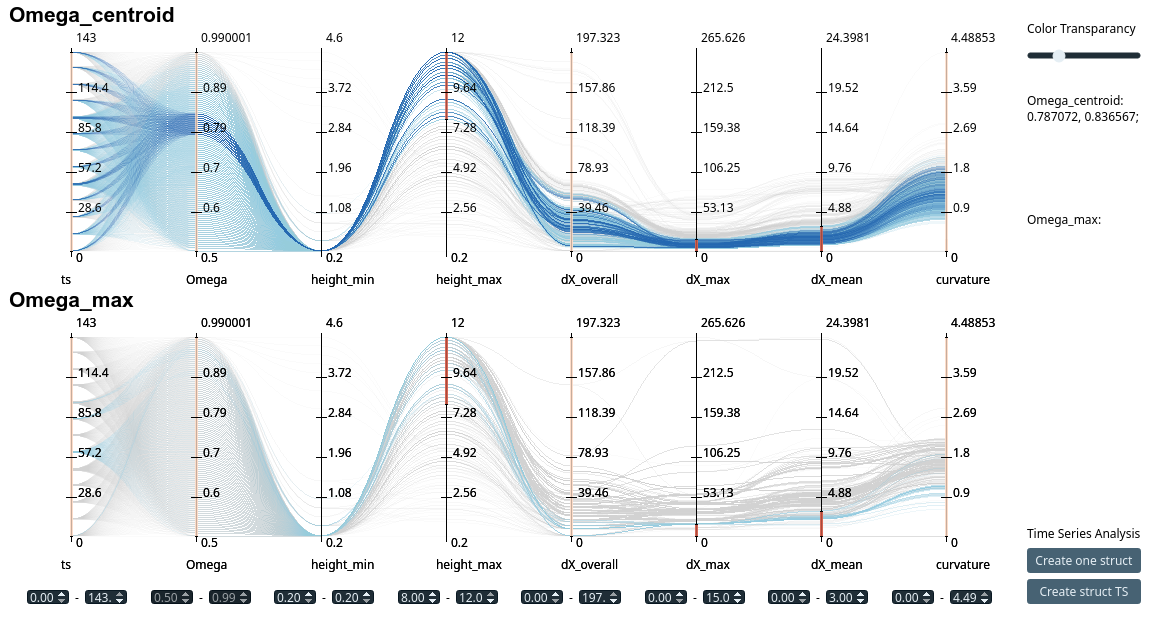}
		\caption{\label{fig:pc_plot_Omega}
			Comparing two TC core line extraction methods using parallel coordinates plot.}
	\end{figure} 
	
\begin{figure}[tb]
		\centering
		\includegraphics[width=0.6\linewidth]{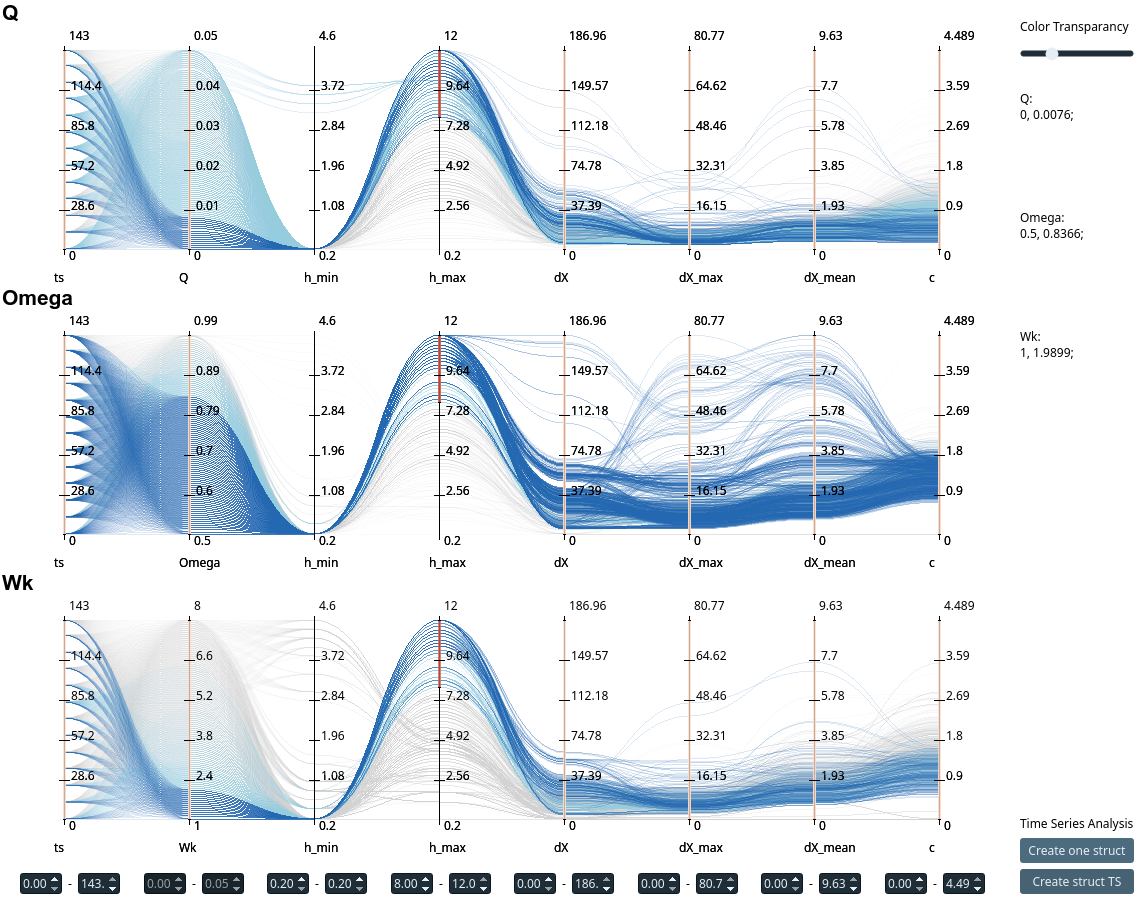}\\
		\includegraphics[width=0.6\linewidth]{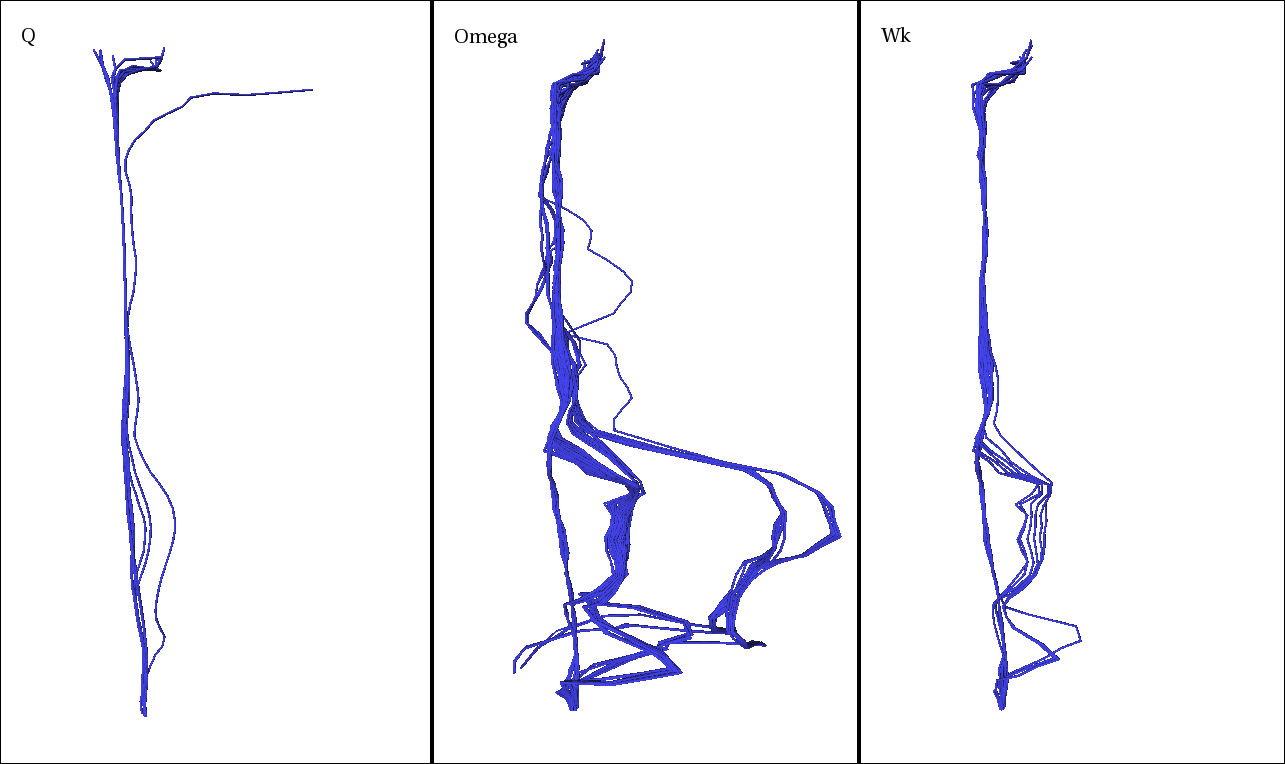}
		\caption{\label{fig:pc_lines_height_restr}
			Parallel coordinates window and core lines identified by $Q$, $\Omega$ and $W_k$ using only height restrictions. }
	\end{figure}

	Next, we focus on the centroid method and utilize the PC tool to investigate the parameter space for three selected indicator quantities.
	First, we applied the height restrictions $h_{min}=\qty{0.2}{\kilo\meter}$ and $h_{max}>\qty{8}{\kilo\meter}$ to ensure that derived lines have a desired height.
	Extracted structures still have a high variation (especially the ones found with $\Omega$), see Fig.~\ref{fig:pc_lines_height_restr}.
	Taking horizontal displacement parameters into consideration, i.e., $\overline{dX}<\qty{3}{\kilo\meter}$ and $\max dX <\qty{15}{\kilo\meter}$, leads to intervals listed in the following table.
	\vspace{-0.4cm}
	\begin{center}
		\begin{tabular}{ c c c} 
			\hline
			indicator & core region & center line  \\ 
			\hline
			$Q$ & [0.002,0.0076] &  [0.0025,0.03] $\cup$ [0.0045,0.0076]\\ 
			$\Omega$ & [0.7574,0.8316] & [0.7871,0.8366]\\ 
			$W_k$ & [1.4949,1.9899] & [1.6364,1.9192]\\ 
			\hline
		\end{tabular}
	\end{center}
	%\newpage
	Identified intervals for TC vortex core region and TC center line overlap but do not perfectly match.
	Overall the intervals derived for center line definition are more strongly restricted.
	
	Finally, extracted lines for the selected time frame are shown in Fig.~\ref{fig:lines_ts}.
	For all time steps, 9, 11, and 5 lines represent the selected intervals corresponding to $Q$, $\Omega$, and $W_k$, respectively.
	
	As a result we find that possible structure definitions of center lines are given per horizontal slice by $x_\omega=(\Bar{x}_\omega,\Bar{y}_\omega)$ with
	\begin{equation*}
		\Bar{x}_\omega = \frac{\sum\limits_{i\in C_\omega}x_i\omega_i}{\sum\limits_{i\in C_\omega}\omega_i} \text{ and } \Bar{y}_\omega = \dfrac{\sum\limits_{i\in C_\omega}y_i\omega_i}{\sum\limits_{i\in C_\omega}\omega_i},
	\end{equation*}
	where $i$ are the indices of grid points $(x_i, y_i)$ within one of the volumes
	\begin{alignat*}{4}
		C_Q &= \{x \in \mathbb{R}^3\:|\: Q(x) > \tau_Q;&&\tau_Q\in [0.0025,0.03]\cup [0.0045,0.0076]\}, \\
		C_\Omega &= \{x \in \mathbb{R}^3\:|\: \Omega(x)> \tau_\Omega;&&\tau_\Omega\in [0.7871,0.8366]\}, \\
		C_{W_k} &= \{x \in \mathbb{R}^3\:|\:W_k(x)>\tau_{W_k};\,&&\tau_{W_k}\in [1.6364,1.9192]\}.
	\end{alignat*}
	
	\begin{figure}[tb]
		\centering
		\includegraphics[width=0.6\linewidth]{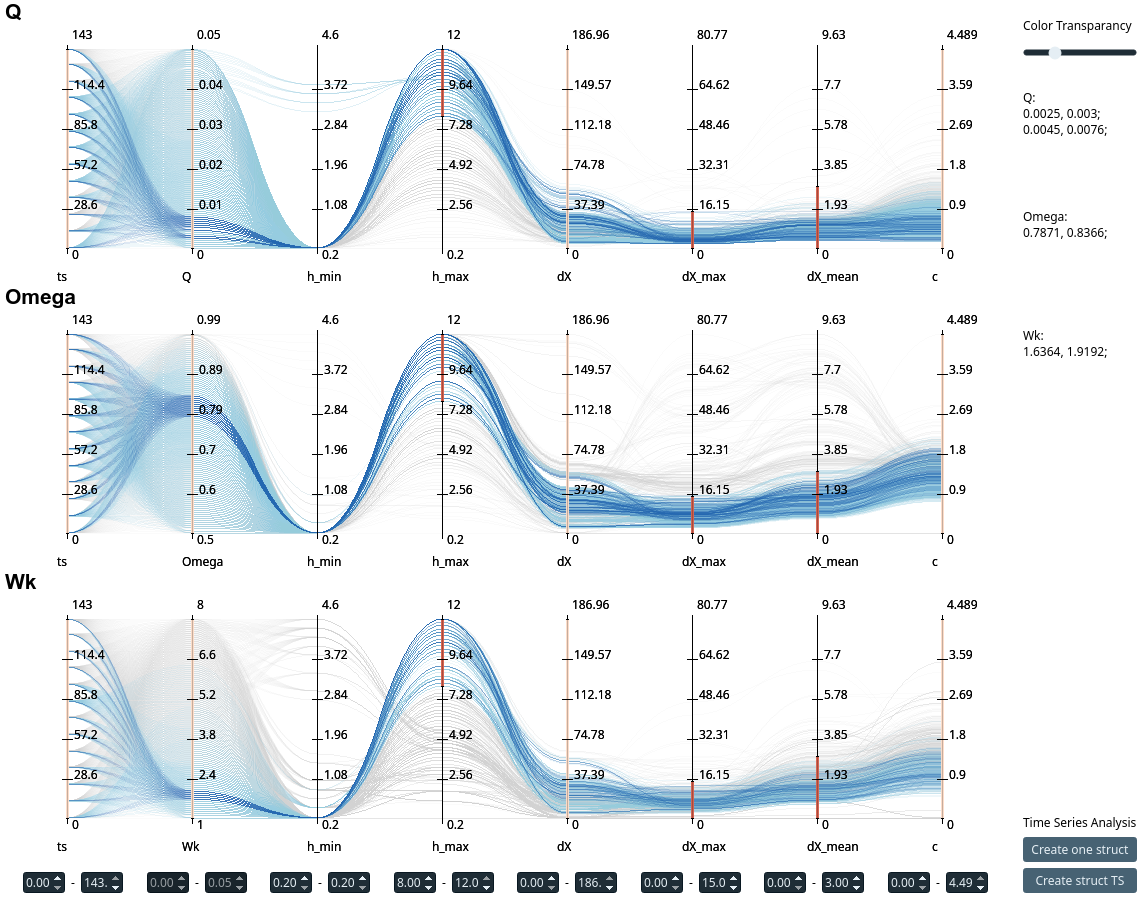}\\
		\includegraphics[width=0.6\linewidth]{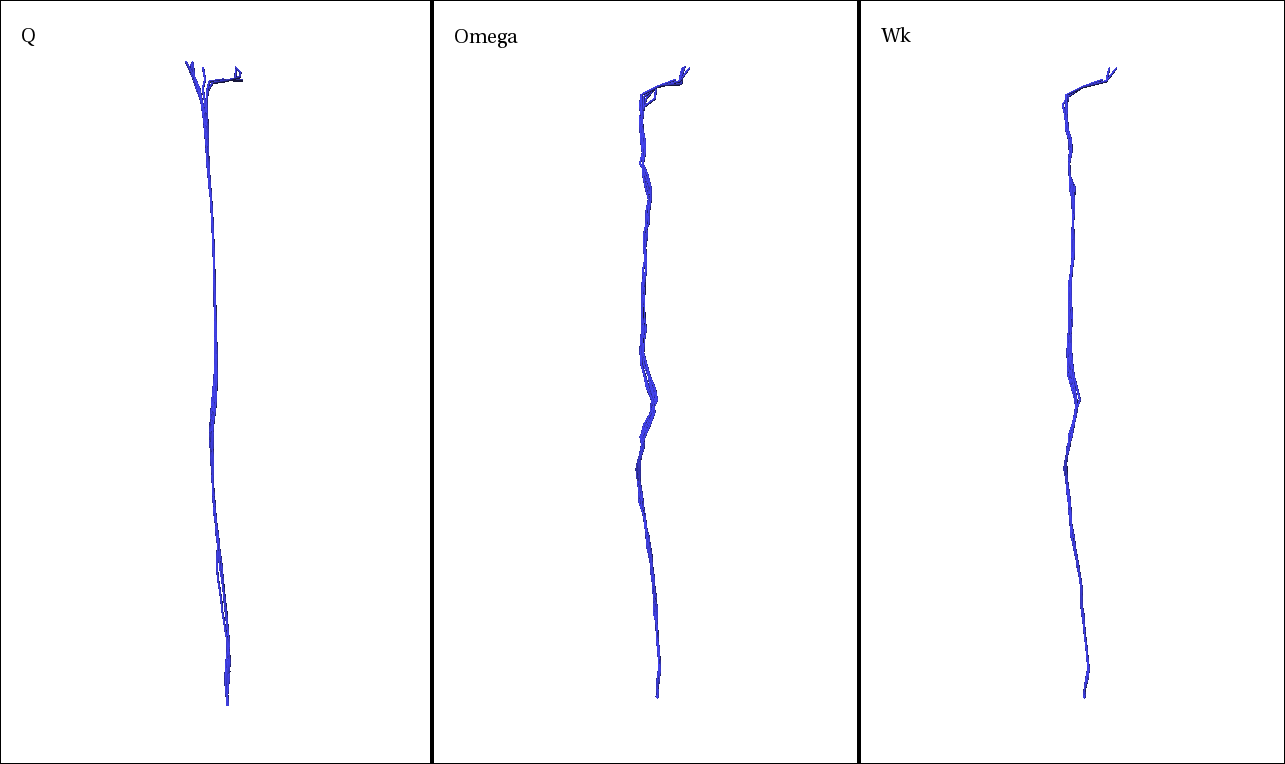}
		\caption{\label{fig:pc_lines_compact}
			Parallel coordinates window and core lines extracted by selecting proper parameter values for three indicator quantities $Q$, $\Omega$ and  $W_k$. }
	\end{figure}
	
	%-------------------------------------------------------------------------
	\section{Discussion and future work}
	\label{sect:discussion}
	
	In many scientific fields, e.g., meteorology, important concepts are described only verbally w.r.t.\ a mental image rather than by precise mathematical definitions.
	Research over the past decades has shown that finding such mathematical descriptions is not an easy task and ``does not just happen''.
	
	%%%
	In this work, we describe a systematic approach to address this problem and demonstrate the applicability of the approach using two structures from meteorology.
	A key question in this approach is, how a mental image can be turned into an precise, formal structure definition.
	This is achieved by identifying and computing suitable attributes that allow one to measure how well an extracted structure matches the mental image, and by subsequently relating these attributes to the complete parameter space of the structure definition including computational methods, indicator quantities, and parameter intervals.
	We use the paradigm of linking \& brushing on the attribute space together with a direct 3D visualization of the selected structures to allow the user to verify whether the obtained structure definition indeed describes the mental image.
	This allows us to narrow the structure definition further and further.
	In order to identify structure definitions that are applicable to all time steps of a time-dependent data set, a common scenario in the analysis of meteorological phenomena, it is essential to look at all time steps at the same time.
	
	%%%    
	One inherent limitation of our approach is that it requires a potentially large number of structures to be extracted for different parameters and time steps along with the computation of all the necessary attributes describing the structures.
	However, this can usually be done automatically in a preprocessing step.
	Once this is done, our approach is interactive.
	
	In order to test the proposed approach, we implemented an interactive prototype tool based on parallel coordinates to identity definitions for two structures from meteorology, vortex core regions and core lines of tropical cyclones (TC). 
	Our aim was not to identify the best method for the extraction of these specific structures.
	Rather, the analyses carried out in this presentation were done to illustrate that it is possible to objectify mental images using interactive exploration by coupling structure attributes and parameter intervals.
	
	One motivation for the development of the presented tool is the fact that the vortex core line is a major ingredient for the asymptotic theory on strongly tilted TCs~\citep{paschke2012motion,dorffel2021dynamics}.
	The TC core line prescribes the origin of a tilted cylindrical coordinate system and both, symmetric and asymmetric structures, are assessed relative to this origin.
	Hence, further evaluations of this theory require a robust definition of the TC core line for which this work not only provides a definition but also an estimate on the uncertainty comprised in the ensemble.
	Along this line, we will apply our prototype tool to further TC data, e.g., Hurricanes Isabel 2003, Earl 2010, Eduoard 2014, and Fiona 2016.
	The question we will ask is whether we can find similar parameter ranges across different TC data including higher-resolution simulations like ICON.
	
	%%%
	Other future work will include the application and extension of our tool to further meteorological structures including convective cells and cold fronts.
	This will show limitations of our current prototype and allow us to generalize the tool further.
	
	%%%
	In the examples we have given, only a one-dimensional parameter space was explored.
	However, the approach and the presented tool also support multi-dimensional parameter spaces.
	The main problem here will be the sampling of the multi-dimensional parameter space and the handling of the potentially very larger number of extracted sample structures.
	Therefore, more elaborate sampling methods than the one used here might be needed.
	
	%%%
	Our prototype implementation implicitly concatenates the attribute selection using the logical AND operation, thus creating a single high-dimensional attribute interval.
	However, the support of more complicated logical expressions might be needed for other structures.
	For this, the development of an expression editor alongside the parallel coordinates plot will be explored.

	\section*{Acknowledgments}
	Natalia Mikula and Tom D\"orffel have been supported by Deutsche Forschungsgemeinschaft (DFG) through grant CRC 1114 ``Scaling Cascades in Complex Systems'',  Project Number 235221301, Project (C06) ``Multi-scale structure of atmospheric vortices''.

	%-------------------------------------------------------------------------
	% bibtex
	\bibliography{bibliography}
	
	% biblatex with biber
	% \printbibliography                
	
	%-------------------------------------------------------------------------
	
\end{document}